\documentclass{aastex62}

\received{June 17, 2019}
\revised{November 14, 2019}
\accepted{November 14, 2019}
\submitjournal{AJ}

\shorttitle{Subaru Polarization Imaging of Misaligned Disks}
\shortauthors{Mayama et al.}

\begin{document}

\title{SUBARU NEAR-INFRARED IMAGING POLARIMETRY OF MISALIGNED DISKS AROUND THE SR24 HIERARCHICAL TRIPLE SYSTEM\footnote{Based on data collected at the Subaru Telescope, which is operated
    by the National Astronomical Observatory of Japan.}}

\correspondingauthor{Satoshi MAYAMA}
\email{mayama\_satoshi@soken.ac.jp}
\author{Satoshi MAYAMA}
\affiliation{The Graduate University for Advanced Studies, SOKENDAI, Shonan Village, Hayama, Kanagawa 240-0193, Japan}

\author{Sebasti\'an P\'EREZ}
\affiliation{Universidad de Santiago de Chile, Av. Libertador Bernardo O'Higgins 3363, Estaci\'on Central, Santiago, Chile}

\author{Nobuhiko KUSAKABE}
\affiliation{Astrobiology Center, NINS, 2-21-1, Osawa, Mitaka, Tokyo 181-8588, Japan}
\affiliation{National Astronomical Observatory of Japan (NAOJ), National Institutes of Natural Sciences (NINS), 2-21-1 Osawa, Mitaka, Tokyo 181-8588,Japan}

\author{Takayuki MUTO}
\affiliation{Division of Liberal Arts, Kogakuin
University, 1-24-2, Nishi-Shinjuku, Shinjuku-ku, Tokyo, 163-8677, Japan}

\author{Takashi TSUKAGOSHI}
\affiliation{National Astronomical Observatory of Japan (NAOJ), National Institutes of Natural Sciences (NINS), 2-21-1 Osawa, Mitaka, Tokyo 181-8588,Japan}

\author{Michael L. Sitko}
\affiliation{Center for Extrasolar Planetary Systems, Space Science Institute, 1120 Paxton Ave., Cincinnati, OH 45208, USA}

\author{Michihiro TAKAMI}
\affiliation{Institute of Astronomy and Astrophysics, Academia Sinica, P.O. Box 23-141, Taipei 10617, Taiwan}

\author{Jun HASHIMOTO}
\affiliation{Astrobiology Center, NINS, 2-21-1, Osawa, Mitaka, Tokyo 181-8588, Japan}
\affiliation{National Astronomical Observatory of Japan (NAOJ), National Institutes of Natural Sciences (NINS), 2-21-1 Osawa, Mitaka, Tokyo 181-8588,Japan}

\author{Ruobing DONG}
\affiliation{Department of Physics \& Astronomy, University of Victoria, 3800 Finnerty Rd, Victoria, BC V8P 5C2, Canada}

\author{Jungmi KWON}
\affiliation{Department of Astronomy, The University of Tokyo, 7-3-1 Hongo, Bunkyo-ku, Tokyo 113-0033, Japan}

\author{Saeko S. HAYASHI}
\affiliation{Department of Astronomical Science, SOKENDAI (The Graduate University for Advanced Studies), 2-21-1 Osawa, Mitaka, Tokyo 181-8588, Japan}
\affiliation{National Astronomical Observatory of Japan (NAOJ), National Institutes of Natural Sciences (NINS), 2-21-1 Osawa, Mitaka, Tokyo 181-8588,Japan}

\author{Tomoyuki KUDO}
\affiliation{Subaru Telescope, NAOJ, NINS, 650 North A'ohoku Place, Hilo, HI 96720, USA}

\author{Masayuki KUZUHARA}
\affiliation{Astrobiology Center, NINS, 2-21-1, Osawa, Mitaka, Tokyo 181-8588, Japan}
\affiliation{National Astronomical Observatory of Japan (NAOJ), National Institutes of Natural Sciences (NINS), 2-21-1 Osawa, Mitaka, Tokyo 181-8588,Japan}

\author{Katherine FOLLETTE}
\affiliation{Amherst College Department of Physics and Astronomy, AC\#2244, PO Box 5000, Merrill Science Center, 15 Mead Drive, Amherst, MA 01002-5000, USA}

\author{Misato FUKAGAWA}
\affiliation{National Astronomical Observatory of Japan (NAOJ), National Institutes of Natural Sciences (NINS), 2-21-1 Osawa, Mitaka, Tokyo 181-8588,Japan}

\author{Munetake MOMOSE}
\affiliation{College of Science, Ibaraki University, 2-1-1 Bunkyo, Mito 310-8512, Japan}

\author{Daehyeon OH}
\affiliation{National Meteorological Satellite Center, 64-18 Guam-gil, Gwanghyewon-myeon, Jincheon-gun, Chungbuk, South Korea}

\author{Jerome DE LEON}
\affiliation{Department of Astronomy, The University of Tokyo,
Hongo 7-3-1, Bunkyo-ku, Tokyo 113-0033, Japan}

\author{Eiji AKIYAMA}
\affiliation{Institute for the Advancement of Higher Education, Hokkaido University, Kita 17, Nishi 8, Kita-ku, Sapporo, Hokkaido, 060-0817, Japan}

\author{John P. WISNIEWSKI}
\affiliation{H. L. Dodge Department of Physics \& Astronomy, University of Oklahoma, 440 W Brooks St., Norman, OK 73019, USA}

\author{Yi YANG}
\affiliation{Astrobiology Center, NINS, 2-21-1, Osawa, Mitaka, Tokyo 181-8588, Japan}
\affiliation{National Astronomical Observatory of Japan (NAOJ), National Institutes of Natural Sciences (NINS), 2-21-1 Osawa, Mitaka, Tokyo 181-8588,Japan}

\author{Lyu ABE}
\affiliation{Laboratoire Lagrange (UMR 7293), Universite de Nice-Sophia Antipolis, CNRS, Observatoire de la Coted'azur, 28 avenue Valrose, 06108 Nice Cedex 2, France}

\author{Wolfgang BRANDNER}
\affiliation{Max Planck Institute for Astronomy, K{\"o}nigstuhl 17, 69117 Heidelberg, Germany}

\author{Timothy D. BRANDT}
\affiliation{Department of Physics, Broida Hall, University of California, Santa Barbara, CA 93106-9530}

\author{Michael BONNEFOY}
\affiliation{Univ. Grenoble Alpes, CNRS, IPAG, F-38000 Grenoble, France}

\author{Joseph C. CARSON}
\affiliation{Department of Physics and Astronomy, College of Charleston, 58 Coming St., Charleston, SC 29424, USA}

\author{Jeffrey CHILCOTE}
\affiliation{Department of Physics, University of Notre Dame, 225 Nieuwland Science Hall, Notre Dame, IN 46556, USA}

\author{Thayne CURRIE}
\affiliation{Subaru Telescope, NAOJ, NINS, 650 North A'ohoku Place, Hilo, HI 96720, USA}


\author{Markus FELDT}
\affiliation{Max Planck Institute for Astronomy, K{\"o}nigstuhl 17, 69117 Heidelberg, Germany}

\author{Miwa GOTO}
\affiliation{Universit{\"a}ts-Sternwarte M{\"u}nchen, Ludwig-Maximilians-Universit{\"a}t, Scheinerstr. 1, 81679 M{\"u}nchen, Germany}

\author{Carol A. GRADY}
\affiliation{Exoplanets and Stellar Astrophysics Laboratory, Code 667, Goddard Space Flight Center, Greenbelt, MD, 20771, USA}
\affiliation{Eureka Scientific, 2452 Delmer, Suite 100, Oakland CA 96002, USA}
\affiliation{Goddard Center for Astrobiology, NASA Goddard Space Flight Center, Greenbelt, MD 20771, USA}

\author{Tyler GROFF}
\affiliation{NASA-Goddard Space Flight Center, Greenbelt, MD, USA}

\author{Olivier GUYON}
\affiliation{Astrobiology Center, NINS, 2-21-1, Osawa, Mitaka, Tokyo 181-8588, Japan}
\affiliation{Subaru Telescope, NAOJ, NINS, 650 North A'ohoku Place, Hilo, HI 96720, USA}
\affiliation{Steward Observatory, University of Arizona, 933 N Cherry Ave., Tucson AZ 85719, USA}

\author{Yutaka HAYANO}
\affiliation{Department of Astronomical Science, SOKENDAI (The Graduate University for Advanced Studies), 2-21-1 Osawa, Mitaka, Tokyo 181-8588, Japan}
\affiliation{National Astronomical Observatory of Japan (NAOJ), National Institutes of Natural Sciences (NINS), 2-21-1 Osawa, Mitaka, Tokyo 181-8588,Japan}
\affiliation{Subaru Telescope, NAOJ, NINS, 650 North A'ohoku Place, Hilo, HI 96720, USA}

\author{Masahiko HAYASHI}
\affiliation{JSPS Bonn Office, Wissenschaftszentrum, Ahrstrasse 58, 53175 Bonn, Germany}

\author{Thomas HENNING}
\affiliation{Max Planck Institute for Astronomy, K{\"o}nigstuhl 17, 69117 Heidelberg, Germany}

\author{Klaus W. HODAPP}
\affiliation{Institute for Astronomy, University of Hawaii, 640 N. A'ohoku Place, Hilo, HI 96720, USA}

\author{Miki ISHII}
\affiliation{National Astronomical Observatory of Japan (NAOJ), National Institutes of Natural Sciences (NINS), 2-21-1 Osawa, Mitaka, Tokyo 181-8588,Japan}

\author{Masanori IYE}
\affiliation{National Astronomical Observatory of Japan (NAOJ), National Institutes of Natural Sciences (NINS), 2-21-1 Osawa, Mitaka, Tokyo 181-8588,Japan}

\author{Markus JANSON}
\affiliation{Department of Astronomy, Stockholm University, AlbaNova University Center, SE-106 91 Stockholm, Sweden}

\author{Nemanja JOVANOVIC}
\affiliation{Jet Propulsion Laboratory, California Institute of Technology, M/S 171-113 4800 Oak Grove Drive Pasadena, CA 91109 USA}

\author{Ryo KANDORI}
\affiliation{Astrobiology Center, NINS, 2-21-1, Osawa, Mitaka, Tokyo 181-8588, Japan}

\author{Jeremy KASDIN}
\affiliation{Department of Astrophysical Science, Princeton University, Peyton Hall, Ivy Lane, Princeton, NJ 08544, USA}

\author{Gillian R. KNAPP}
\affiliation{Department of Astrophysical Science, Princeton University, Peyton Hall, Ivy Lane, Princeton, NJ 08544, USA}


\author{Julien LOZI}
\affiliation{Subaru Telescope, NAOJ, NINS, 650 North A'ohoku Place, Hilo, HI 96720, USA}

\author{Frantz MARTINACHE}
\affiliation{ Universit\'e C\^ote d'Azur, Observatoire de la C\^ote d'Azur, CNRS, Laboratoire Lagrange, France}

\author{Taro MATSUO}
\affiliation{Department of Earth and Space Science, Graduate School of Science, Osaka University, 1-1 Machikaneyamacho, Toyonaka, Osaka 560-0043, Japan}

\author{Michael W. MCELWAIN}
\affiliation{Exoplanets and Stellar Astrophysics Laboratory, Code 667, Goddard Space Flight Center, Greenbelt, MD, 20771, USA}

\author{Shoken MIYAMA}
\affiliation{Hiroshima University, 1-3-2 Kagamiyama, Higashihiroshima, Hiroshima 739-8511, Japan}

\author{Jun-Ichi MORINO}
\affiliation{National Astronomical Observatory of Japan (NAOJ), National Institutes of Natural Sciences (NINS), 2-21-1 Osawa, Mitaka, Tokyo 181-8588,Japan}

\author{Amaya MORO-MARTIN}
\affiliation{Space Telescope Science Institute(STScI), 3700 San Martin Drive, Baltimore, MD 21218}

\author{Takao Nakagawa}
\affiliation{Institute of Space and Astronautical Science, Japan Aerospace Exploration Agency, 3-1-1 Yoshinodai, Chuo-ku, Sagamihara, Kanagawa 252-5210, Japan}

\author{Tetsuo NISHIMURA}
\affiliation{Subaru Telescope, NAOJ, NINS, 650 North A'ohoku Place, Hilo, HI 96720, USA}

\author{Tae-Soo PYO}
\affiliation{Department of Astronomical Science, SOKENDAI (The Graduate University for Advanced Studies), 2-21-1 Osawa, Mitaka, Tokyo 181-8588, Japan}
\affiliation{Subaru Telescope, NAOJ, NINS, 650 North A'ohoku Place, Hilo, HI 96720, USA}

\author{Evan A. RICH}
\affiliation{H. L. Dodge Department of Physics \& Astronomy, University of Oklahoma, 440 W Brooks St., Norman, OK 73019, USA}

\author{Eugene SERABYN}
\affiliation{Jet Propulsion Laboratory, California Institute of Technology, M/S 183-900 4800 Oak Grove Drive Pasadena, CA 91109, USA}

\author{Hiroshi SUTO}
\affiliation{Astrobiology Center, NINS, 2-21-1, Osawa, Mitaka, Tokyo 181-8588, Japan}
\affiliation{National Astronomical Observatory of Japan (NAOJ), National Institutes of Natural Sciences (NINS), 2-21-1 Osawa, Mitaka, Tokyo 181-8588,Japan}

\author{Ryuji SUZUKI}
\affiliation{National Astronomical Observatory of Japan (NAOJ), National Institutes of Natural Sciences (NINS), 2-21-1 Osawa, Mitaka, Tokyo 181-8588,Japan}

\author{Naruhisa TAKATO}
\affiliation{Department of Astronomical Science, SOKENDAI (The Graduate University for Advanced Studies), 2-21-1 Osawa, Mitaka, Tokyo 181-8588, Japan}
\affiliation{Subaru Telescope, NAOJ, NINS, 650 North A'ohoku Place, Hilo, HI 96720, USA}

\author{Hiroshi TERADA}
\affiliation{National Astronomical Observatory of Japan (NAOJ), National Institutes of Natural Sciences (NINS), 2-21-1 Osawa, Mitaka, Tokyo 181-8588,Japan}

\author{Christian THALMANN}
\affiliation{Swiss Federal Institute of Technology (ETH Zurich), Institute for Astronomy, Wolfgang-Pauli-Strasse 27, CH-8093 Zurich, Switzerland}

\author{Daigo TOMONO}
\affiliation{Subaru Telescope, NAOJ, NINS, 650 North A'ohoku Place, Hilo, HI 96720, USA}

\author{Edwin L. TURNER}
\affiliation{Department of Astrophysical Science, Princeton University, Peyton Hall, Ivy Lane, Princeton, NJ 08544, USA}
\affiliation{Kavli Institute for Physics and Mathematics of the Universe, The University of Tokyo, 5-1-5 Kashiwanoha, Kashiwa, Chiba 277-8568, Japan}

\author{Makoto WATANABE}
\affiliation{Department of Cosmosciences, Hokkaido University, Kita-ku, Sapporo, Hokkaido 060-0810, Japan}

\author{Toru YAMADA}
\affiliation{Institute of Space and Astronautical Science, Japan Aerospace Exploration Agency, 3-1-1 Yoshinodai, Chuo-ku, Sagamihara, Kanagawa 252-5210, Japan}

\author{Hideki TAKAMI}
\affiliation{Department of Astronomical Science, SOKENDAI (The Graduate University for Advanced Studies), 2-21-1 Osawa, Mitaka, Tokyo 181-8588, Japan}
\affiliation{National Astronomical Observatory of Japan (NAOJ), National Institutes of Natural Sciences (NINS), 2-21-1 Osawa, Mitaka, Tokyo 181-8588,Japan}

\author{Tomonori USUDA}
\affiliation{Department of Astronomical Science, SOKENDAI (The Graduate University for Advanced Studies), 2-21-1 Osawa, Mitaka, Tokyo 181-8588, Japan}
\affiliation{National Astronomical Observatory of Japan (NAOJ), National Institutes of Natural Sciences (NINS), 2-21-1 Osawa, Mitaka, Tokyo 181-8588,Japan}

\author{Taichi UYAMA}
\affiliation{Department of Astronomy, The University of Tokyo, 7-3-1 Hongo, Bunkyo-ku, Tokyo 113-0033, Japan}

\author{Motohide TAMURA}
\affiliation{Department of Astronomy, The University of Tokyo, 7-3-1 Hongo, Bunkyo-ku, Tokyo 113-0033, Japan}
\affiliation{Astrobiology Center, NINS, 2-21-1, Osawa, Mitaka, Tokyo 181-8588, Japan}
\affiliation{National Astronomical Observatory of Japan (NAOJ), National Institutes of Natural Sciences (NINS), 2-21-1 Osawa, Mitaka, Tokyo 181-8588,Japan}

\begin{abstract}
The SR24 multi-star system hosts both circumprimary and circumsecondary disks, which are strongly misaligned with each other. The circumsecondary disk is circumbinary in nature. Interestingly, both disks are interacting, and they possibly rotate in opposite directions. To investigate the nature of this unique twin disk system, we present 0\rlap.~{$''$}1 resolution near-infrared polarized intensity images of the circumstellar structures around SR24, obtained with HiCIAO mounted on the Subaru 8.2 m telescope.
Both the circumprimary disk and the circumsecondary disk are resolved and have elongated features. While the position angle of the major axis and radius of the near-IR(NIR) polarization disk around SR24S are 55~$^{\circ}$ and 137 au, respectively, those around SR24N are 110~$^{\circ}$ and 34 au, respectively. With regard to overall morphology, the circumprimary disk around SR24S shows strong asymmetry, whereas the circumsecondary disk around SR24N shows relatively strong symmetry.  Our NIR observations confirm the previous claim that the circumprimary and circumsecondary disks are misaligned from each other. Both the circumprimary and circumsecondary disks show similar structures in $^{12}$CO observations in terms of its size and elongation direction. This consistency is because both NIR and $^{12}$CO are tracing surface layers of the flared disks. As the radius of the polarization disk around SR24N is roughly consistent with the size of the outer Roche lobe, it is natural to interpret the polarization disk around SR24N as a circumbinary disk surrounding the SR24Nb-Nc system.
\\
\\
\\
\\

\end{abstract}



\section{Introduction}

Observationally, there have been many young binary stars hosting a circumprimary disk misaligned with respect to either a circumsecondary disk, a circumbinary disk, or a binary orbital plane (e.g., HK Tau, \citet{Jensen2014}, L1551 NE, \citet{Takakuwa2017}, GG Tau, \citet{Aly2018}, IRS43, \citet{Brinch2016}, GW Ori, \citet{Czekala2017}, HD98800, \citet{Kennedy2019}). These circumprimary and circumsecondary disks are directly imaged as two single disks. More recently, another type of young misaligned disks is beginning to be observed. They are misaligned inner disks with respect to outer disks both surrounding single transitional object (e.g., HD142527, \citet{Casassus2015}, HD100453, \citet{Benisty2017, Vanderplas2019}, HD143006, \citet{Benisty2018}, HD135344B, \citet{Stolker2016}, DoAr 44, \citet{Casassus2018}, J1604, \citet{Mayama2012}; \citet{Mayama2018}). Even in the earlier stage of protostar evolution, a warped disk around a protostar IRAS 04368+2557 was discovered with ALMA\citep{Sakai2019}.


Some promising mechanisms that have been claimed to address theoretically the origin of inner disks misaligned with respect to outer disks are as follows:
1) the rotation axis of the disk system is misaligned with respect to the magnetic field direction (e.g., \citet{2010MNRAS.409L..39C}); 2) anisotropic accretion of gas with different rotational axes (e.g., \citet{Bate2018}); 3) a misaligned massive planet with respect to an outer disk tilting an inner disk (e.g., \citet{Nealon2019}, \citet{Zhu2019}).      
In the third mechanism, the planet is assumed to be sufficiently massive to open a gap in the disk.
Such planets can become misaligned with respect to an outer disk through secular interaction with an external misaligned companion \citep{2016ApJ...817...30L,2016MNRAS.458.4345M}, or through precessional resonances \citep{2017MNRAS.469.2834O}. In both cases, the inner disk (within the planet/companion orbital radius) might become aligned to the orbital plane of the planet, thus becoming misaligned with respect to the outer disk.  


Among observed misaligned disks so far, ALMA observations shed light on SR24, the target of this study, because \citet{2017ApJ...845...10F} suggests that the circumprimary disk is strongly misaligned (108$^{\circ}$) with respect to the circumsecondary disk and both disks possibly rotate in opposite directions as observed from Earth, in projection. Here, the target of this study is introduced.

SR~24, also known as HBC~262, is located in the Ophiuchus star-forming region. GAIA DR2 reported that SR~24 is located at a distance of 114 pc\citep{Gaia2018}.  
 SR~24 is a hierarchical multiple system composed of primary SR~24~S and secondary SR~24~N.
 SR~24~S is classified as a K2 type class~II T Tauri star and has a mass of \textgreater1.4~\textit{M}$_\odot$ \citep{Cohen79, Correia06}. 
 SR~24~N, located 5\rlap.{$''$}2 north at a position angle (PA) of 348$^{\circ}$ \citep{Reipurth93}, is classified as an M0.5 type class~II T Tauri star \citep{Cohen79}.
 \citet{Simon95} observed that SR~24~N itself is a binary system of SR~24~Nb and SR~24~Nc with a projected separation of 0\rlap.{$''$}197.  The eccentricity of the orbit of SR~24~Nb and SR~24~Nc is derived as $0.64_{-0.10}^{+0.13}$\citep{Schaefer2018}.
 The spectral type and mass of SR~24~Nb are K4-M4 and 0.61~\textit{M}$_\odot$, respectively \citep{Correia06}, whereas those of SR~24~Nc are K7-M5 and 0.34~\textit{M}$_\odot$, respectively \citep{Correia06}.

\citet{Nuernberger98} observed the dust emission associated with the SR~24~S, whereas for the SR~24~N, they derived only an upper limit of the flux density based on their 1.3 mm map.  
At 10~$\mu$m, which is an indicator of warm circumstellar dust in the inner part of the disk, both south and north components showed roughly equal emission.  
Thus, the 10~$\mu$m measurements indicate that the inner part of the disk around SR~24~N is still present, whereas the non-detection of 1.3~mm emission from SR~24~N indicates a lack of cold circumstellar dust in the outer part of the disk.  
\citet{Nuernberger98} suggested that this was likely due to enhanced disk accretion or destruction caused by the presence of the SR~24~Nc.

\citet{Andrews05} presented high-resolution aperture synthesis images from the submillimeter array of the 1.3 mm continuum and CO J~=~2-1 line emission from the disks around the components of SR~24.  
In their image, SR~24~S is associated with a circumstellar disk detected both in the continuum and CO line emission with properties typical of those around single T Tauri stars, whereas SR~24~N is only detected in CO line emission and not in the continuum.  
Based on their observations, they suggested that SR~24~N was surrounded by at least one circumstellar disk and a circumbinary gas disk, presumably with a dynamically carved gap.


\citet{Andrews07} presented a high-spatial-resolution submillimeter 1330~$\mu$m continuum image of SR~24 using SMA.
They modeled the circumstellar disk around SR~24~S by using broadband spectral energy distribution and submillimeter visibilities to derive the physical parameters of the disk.
Their results show that the outer radius, inclination, and PA of the circumprimary disk around SR~24~S are 500$^{+500}_{-175}$ AU, 57$^{\circ}$, and 25$^{\circ}$, respectively.

\citet{2017ApJ...845...10F} reported ALMA data and detected 1.3 mm continuum emission from SR24N for the first time in this wavelength domain. The mass associated with the SR24S and SR24N disks is derived as 0.025 \textit{M}$_\odot$ and 4$\times$10$^{-5}$~\textit{M}$_\odot$, respectively.
In addition, their $^{12}$CO(2-1) ALMA and SMA velocity cubes show three main features (i) a gas reservoir extending north-northwest of SR24N, (ii) a bridge of gas connecting SR24N with SR24S disks, and (iii) an elongated and blueshifted feature due southwest of SR24S.

 In the near-infrared (NIR), \citet{Mayama10} resolved both circumprimary and circumsecondary disks around SR24S and SR24N, respectively. Their 0\rlap.{$''$}1 observation detected a bridge of infrared emission connecting the two disks and a long spiral arm extending from the circumprimary disk.  

\citet{Zhang13} conducted H$_2$ NIR imaging observation to search for molecular hydrogen emission line objects. Although their observation covers an area of $\sim$0.11 deg$^2$ toward the L1688 core in the $\rho$ Ophiuchi molecular cloud including the area where SR24 is located, they do not detect any emission from SR24.

 As SR24 is a complex hierarchical triple system, there are still many unanswered questions in this regard. Therefore, in this paper, we present high-resolution NIR polarimetric images of SR24 south and north as data. High-resolution polarimetric imaging is a powerful tool to study the structure of protoplanetary disks. 
The rest of this paper is organized as follows. Observations and data reduction procedures are described in Section 2.
The results and discussion are presented in Sections 3 and 4,
respectively. Section 5 summarizes the conclusions.

\section{Observations and Data Reduction}

We performed polarimetry in the $H$-band (1.6~$\mu$m) toward SR24 
using the high-resolution imaging instrument HiCIAO 
\citep{Tamura06, Hodapp06} with a dual-beam  
polarimeter mounted on the Subaru 8.2~m Telescope on August 2, 2011. These observations are part of the high-contrast imaging survey,
Strategic Explorations of Exoplanets and Disks with Subaru
\citep[SEEDS;][]{Tamura09}. The polarimetric observation mode
acquires \textit{o}-rays 
and \textit{e}-rays simultaneously, and images a field of view of
10$''\times$20$''$ with a pixel  
scale of 9.5~mas/pixel. SR24S was observed without an occulting mask in order to image 
the innermost region around the central star. The exposures were sequentially performed at four position angles (P.A.s) of the half-wave plate, which are PA = 0$^{\circ}$, 45$^{\circ}$, 
22.5$^{\circ}$, and 67.5$^{\circ}$, in one rotation cycle to measure the Stokes parameters. The integration time 
per wave plate position was 15 s and the total integration time of
the polarization intensity (hereafter $PI$)  
image was 1140 s.
The adaptive optics system \citep[AO188;][]{Hayano10} provides a diffraction-
limited and almost stable stellar point spread function (PSF).

The Image Reduction and Analysis Facility software (IRAF\footnote{
    IRAF is distributed by the National Optical Astronomy Observatory, which is operated by the Association of Universities for Research in Astronomy, Inc., under cooperative agreement with the National Science Foundation.}) was used for
data reduction.   
We follow the polarimetric data reduction technique described in
\citet{Hashimoto11} and \citet{Muto12},  
in which the standard approach for polarimetric differential imaging \citep{Hinkley09} was adopted.  
By subtracting two images of extraordinary and ordinary rays at each wave plate position, 
we obtained $+Q$, $-Q$, $+U$, and $-U$ images, from which $2Q$ and $2U$ images were obtained through another subtraction to eliminate the remaining aberration. The $PI$ was then
calculated by $PI =\sqrt{Q^2+U^2}$.  
The instrumental polarization of HiCIAO at the Nasmyth platform was
corrected by following \citet{Joos08}. 



\section{Results}

\subsection{SR24S circumprimary disk} \label{bozomath}

The $H$-band $PI$ image of SR24S after subtracting the polarized halo is presented in Figure \ref{fig1}(a). 
The polarized signal corresponds to stellar light scattered of the surface of small dust particles which are mixed with the circumstellar gas.
Disk inner regions around SR24S have appeared at 0\rlap.{$''$}1.  
The bridge and spiral arm, which were detected in \citet{Mayama10}, are not detected with this observation possibly owing to limited observation time which provided a modest signal-to-noise ratio.
While the CIAO image in \citet{Mayama10} revealed the outer part of the outer disk, the relatively inner part of the outer disk is mainly observed at this time in this $PI$ image with HiCIAO.
The circumstellar structure around SR24S has elongated features both to the northeast and southeast directions.

Along the major axis, $PI$ on the northeast side is 7.6 times stronger than that on the southwest side at around 0\rlap.{$''$}25 from the primary source. Along the minor axis, $PI$ on the southeast side is 3.7 times stronger than that on the northwest side at around 0\rlap.{$''$}5 from the primary source.  These show strong asymmetry along both the major and minor axes.

Figure \ref{fig1}(b) shows $H$-band polarization vectors superposed on the $PI$ image.
Although most of the circumprimary structures around SR24S show a centrosymmetric vector pattern, the north-northwest and southwest circumstellar structures do not show such a pattern.  
Considering the separation of SR~24~N-S, the deviation from centrosymmetric polarization angles is probably because the circumstellar disk around SR24S is partly illuminated also by SR24N.  The illumination from a relatively far star is reported around other young multiple systems. Krist et al. (1998), Hioki et al. (2011), and Gledhill and Scarrott (1989) suggested that the northern portion of FS Tau circumbinary disk is illuminated by Haro 6-5 B located 20'' (2800 AU) west of the FS Tau binary.  Many polarization vectors around the FS Tau binary deviate from the larger centrosymmetric pattern in their maps.
  In addition, the azimuth angles of these two regions at the northwest and southwest of SR~24~S are consistent with the disk regions connecting the north bridge and southwest spiral arm shown in \citet{Mayama10}. Therefore, these undetected bridge and arm structures might induce local polarization structures that deviate from the larger centrosymmetric pattern, disturbing the centrosymmetric polarization vector pattern around SR24S.

Figure \ref{fig7}(a) shows the radial surface brightness profile of SR24S along the major axis.
 In the northeast direction, the surface brightness along the major axis decreases as $r$$^{-1.8}$ from 0\rlap.{$''$}4 to 1\rlap.{$''$}2, and decreases as $r$$^{-1.1}$ from 1\rlap.{$''$}3 to 1\rlap.{$''$}4. In the southwest direction, the surface brightness along the major axis decreases 
as $r$$^{-1.1}$ from 0\rlap.{$''$}4 to 0\rlap.{$''$}6, and decreases 
as $r$$^{-0.3}$ from 0\rlap.{$''$}7 to 0\rlap.{$''$}8.  Figure \ref{fig7}(b)
shows the radial profile of the surface brightness along
the minor axis. 
In the northwest direction, the surface brightness along the minor axis decreases as
$r$$^{-1.7}$ from 0\rlap.{$''$}2 to 0\rlap.{$''$}4, and decreases as $r$$^{-0.02}$ from 0\rlap.{$''$}5 to
0\rlap.{$''$}7. In the southeast direction, the surface brightness along the major axis decreases as
$r$$^{-1.8}$ from 0\rlap.{$''$}2 to 1\rlap.{$''$}0.  
The typical error in the power law index is $\sim$0.1.

The radial profiles in the northeast direction along the major axis show a change of
slope beyond 1\rlap.{$''$}2. Thus, our observations indicate that NIR polarization disk seen in scattered light has a radius of 1\rlap.{$''$}2, while there would possibly be structures which are not illuminated by the central star beyond this NIR polarization radius.  
The derived semi-major axis is called ``NIR polarization radius'' in this paper.
The PA of this NIR circumprimary disk is derived as 55$^{\circ}$ as it is the brightest angle.

\subsection{SR24N circumsecondary disk} \label{bozomath2}

\begin{figure}
\includegraphics[angle=0,scale=1.1]{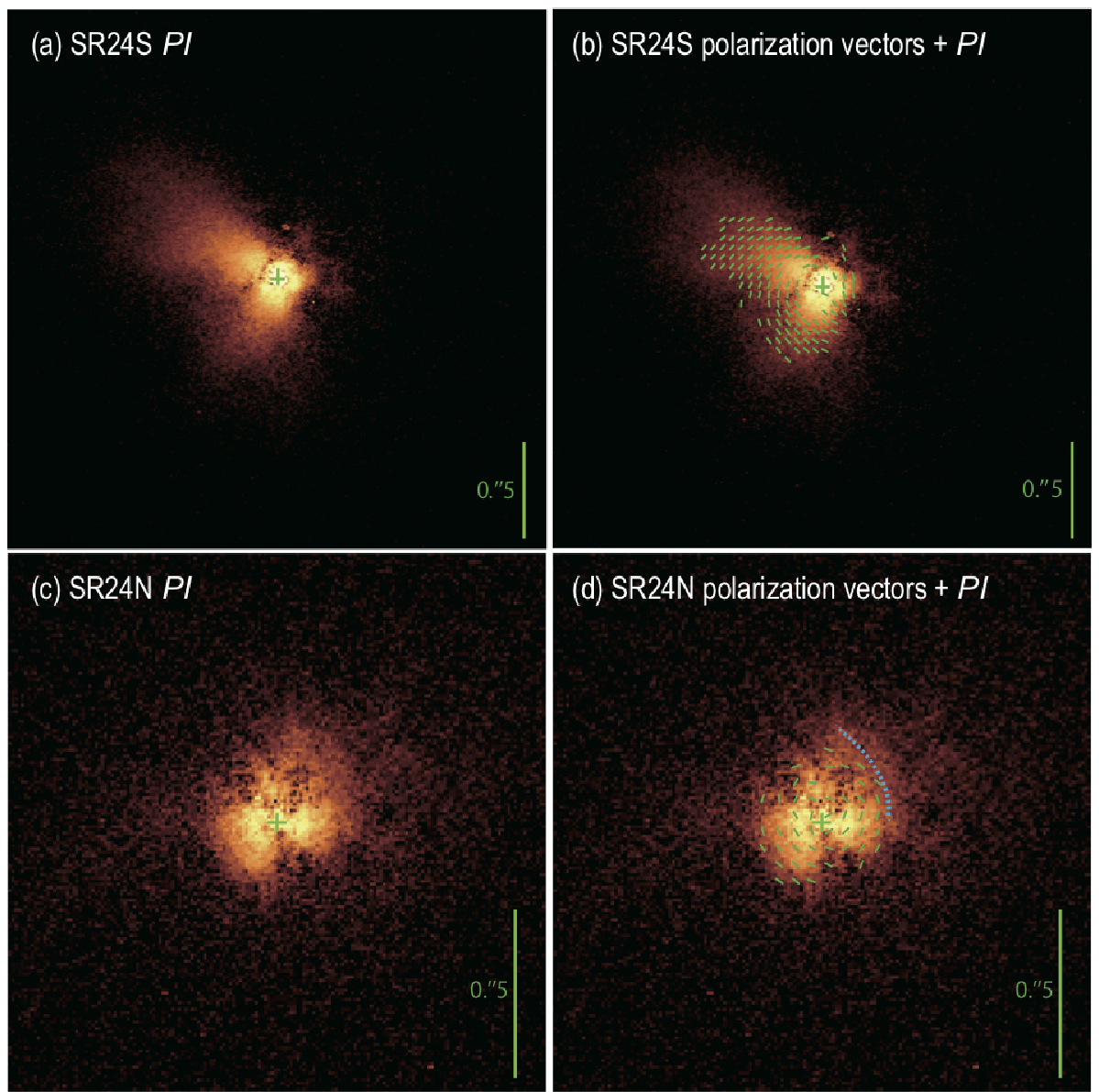}
\caption{$H$-band Subaru+HiCIAO images of SR24. Here, north is up, and east is to the left. The length of the bar indicates $0''.5$ arcsec. The plus sign denotes the position of the central star, SR24S for (a)(b) and SR24N for (c)(d). SR24Nb and SR24Nc are not separately plotted because they are not resolved.
(a) $PI$ image of SR24S.  
The field of view~(FOV) is $2''.8 \times 2''.8$.
(b) $H$-band polarization vectors superposed on the $PI$ image of SR24S.
The vector directions indicate the angles of
polarization.  
The vector's lengths are arbitrary. 
The FOV is $2''.8 \times 2''.8$.  
(c) $PI$ image of SR24N. The FOV is $1''.6 \times 1''.6$.  
(d) $H$-band polarization vectors superposed on the $PI$ image of SR24N.
The vector directions indicate the angles of
polarization.  
The vector's lengths are arbitrary. 
The FOV is $1''.6 \times 1''.6$.
}
\label{fig1}
\end{figure}

Figure \ref{fig1}(c) shows the $H$-band $PI$ image of SR24N after subtracting the polarized halo.  
Figure \ref{fig1}(d) shows the polarization vectors overlaid on the $PI$ image of SR24N. SR24Nb-Nc is not spatially resolved with our Subaru observations. This is because our 0\rlap.{$''$}1 resolution is not sufficiently high enough to resolve SR24Nb-Nc. Based on an orbit calculated by \citet{Schaefer2018}, the separation between SR24Nb-Nc at the time of our observations in 2011 should be much smaller than 93.73$\pm$1.58 mas, which was the closest in time to our Subaru observations and observed by Keck in 2014.  In this paper, we consider SR24Nb and SR24Nc together as SR24N and plot SR24N with a green plus sign in Figure \ref{fig1}(c) and (d). 

\begin{figure}
\includegraphics[angle=0,scale=0.95]{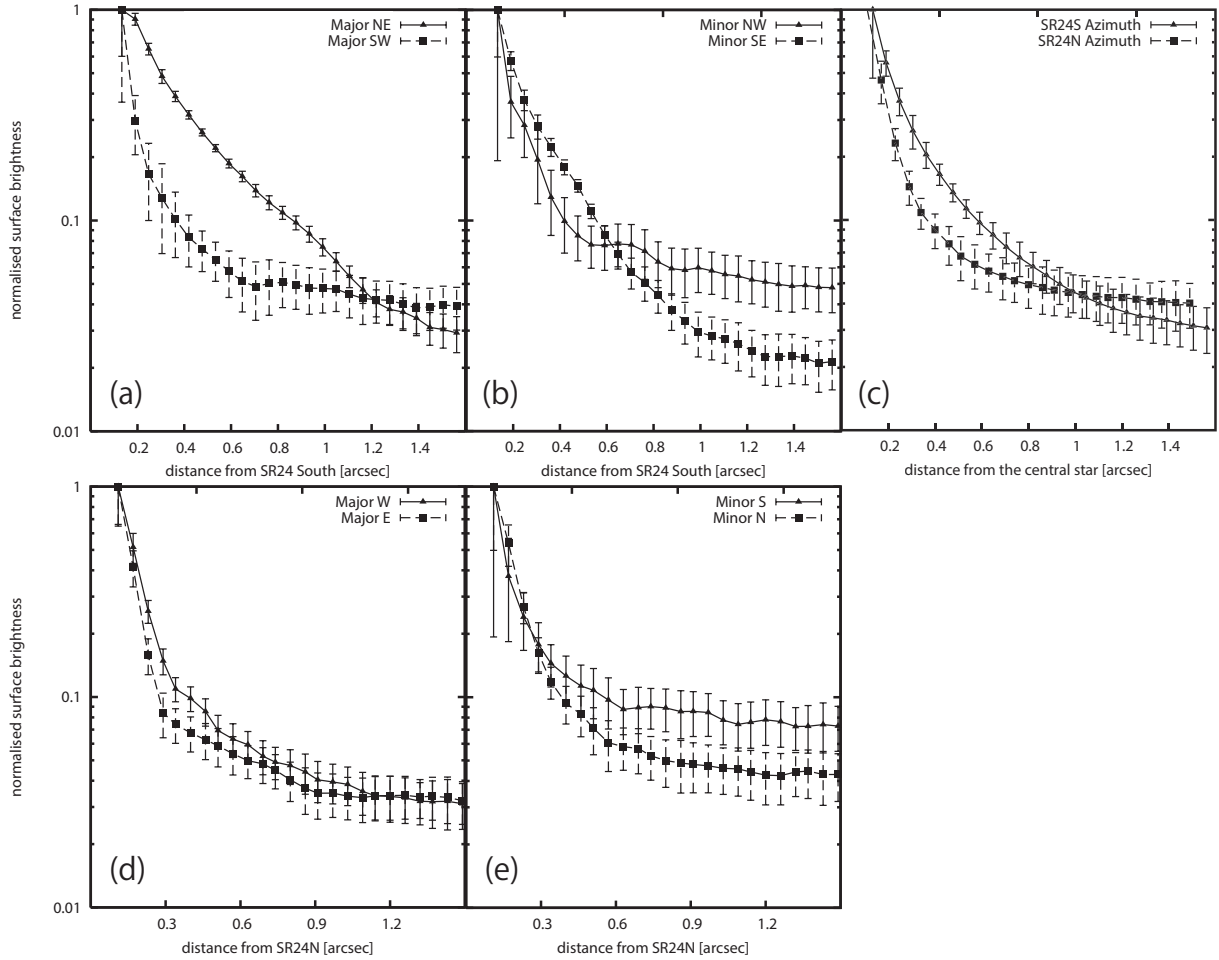}
\caption{Radial surface brightness profiles of SR24 south and SR24 north. (a) The radial surface brightness profile of the primary source SR24 south along the major axis. NE and SW radial profiles are averaged over 45$^{\circ}$~\textless PA \textless ~65$^{\circ}$
 and 225$^{\circ}$~\textless PA \textless ~245$^{\circ}$,
 respectively. (b) Radial surface brightness profile of the primary source SR24 south along the minor axis.  Azimuth radial profile is also displayed.  NW and SE radial profiles are averaged over 315$^{\circ}$~\textless PA \textless ~335$^{\circ}$
 and 135$^{\circ}$~\textless PA \textless ~155$^{\circ}$,
 respectively. (c) Azimuthally averaged normalized surface brightness of SR24 south and SR24 north. (d) Radial surface brightness profile of the secondary source SR24 north along the major axis. West and east radial surface brightness profiles are averaged over 275$^{\circ}$~\textless PA \textless ~295$^{\circ}$
 and 95$^{\circ}$~\textless PA \textless ~115$^{\circ}$,
 respectively. (e) Radial surface brightness profile of the secondary source SR24 north along the minor axis.  Azimuth radial profile is also displayed.  South and north radial profiles are averaged over 185$^{\circ}$~\textless PA \textless ~205$^{\circ}$
 and 5$^{\circ}$~\textless PA \textless ~25$^{\circ}$,
 respectively. 
} 
\label{fig7}
\end{figure}

All the circumsecondary structures around SR24N show a centrosymmetric vector pattern in contrast to SR24S. There are elongated emissions in the east-west direction. This elongated direction is nearly consistent with the CIAO observations. Figure \ref{fig7}(d) shows the radial surface brightness profile of SR24N along the major axis.
 The error bars shown in Figure \ref{fig7} represent the calculated standard deviation.
 In the west direction, the surface brightness along the major axis decreases as $r$$^{-2.1}$ from 0\rlap.{$''$}1 to 0\rlap.{$''$}3, and decreases as $r$$^{-1.0}$ from 0\rlap.{$''$}3 to 0\rlap.{$''$}8. In the east direction, the surface brightness along the major axis decreases 
as $r$$^{-2.6}$ from 0\rlap.{$''$}1 to 0\rlap.{$''$}3, and decreases 
as $r$$^{-0.7}$ from 0\rlap.{$''$}3 to 0\rlap.{$''$}8.  Figure \ref{fig7}(e)
shows the radial profile of the surface brightness along
the minor axis. 
In the south direction, the surface brightness along the minor axis decreases as
$r$$^{-1.6}$ from 0\rlap.{$''$}1 to 0\rlap.{$''$}3. 
In the north direction, the surface brightness along the minor axis decreases as
$r$$^{-2.0}$ from 0\rlap.{$''$}1 to 0\rlap.{$''$}3.  

The radial profiles in the east and west directions along the major axis show a change of
slope beyond 0\rlap.{$''$}3. Thus, our observations indicate that the NIR polarization disk seen in scattered light has a radius of 0\rlap.{$''$}3, while there would possibly be structures which are not illuminated by central star beyond this NIR polarization radius.  
The PA of the circumsecondary disk is derived as 110$^{\circ}$.

Figure \ref{fig7}(c) shows the azimuth-averaged radial surface brightness profile of SR24S and SR24N.
The surface brightness of SR24S decreases as $r$$^{-1.5}$ from 0\rlap.{$''$}2 to 1\rlap.{$''$}0.
The surface brightness of SR24N decreases as $r$$^{-2.1}$ from 0\rlap.{$''$}1 to 0\rlap.{$''$}3.
The typical uncertainty of the measured power law index is $\sim$0.1.
As shown in Figure \ref{fig7}(c), the azimuth radial surface brightness of SR24N has a steeper profile than that of SR24S.
Our observations also show that the SR24S disk is more spatially extended than the SR24N disk.


\section{Discussion}

\subsection{Circumbinary disk surrounding SR24 Nb-Nc}
There appears to be a marginal detection of an arc-shaped structure emanating from the SR24N circumsecondary disk as indicated with blue dashed line in Figure \ref{fig1}(d). It begins at the west side of the SR24N disk, extending north first, then curving to the northeast.  
The polarization vectors in the region of this arc structure face the central star SR24N, indicating that this arc is not an artifact but a real structure illuminated by the central star and is physically connected to the outer edge of the circumsecondary disk associated with SR24N.
As this morphology is symmetric to the bridge emanating from the east side of the SR24N disk also observed using both CIAO and HST, this morphology might be attributed to binary formation.

Adopting the separation between Nb-Nc to be 0."16 as measured by HST observations and the mass ratio, q, of 0.56 based on \citet{Correia06}, the size of the outer Roche lobe and the distance from SR24Nb to L2 point are derived as 0."31 and 0."26 in radius, respectively. As the measured radius 0."3 of the polarization disk around SR24N is roughly consistent with the computed size of the outer Roche lobe, it is natural to interpret the polarization disk around SR24N detected with HiCIAO as a circumbinary disk surrounding the SR24Nb-Nc system. The measured average distance to the arc-shaped structure is 0."26  and it is almost consistent with the computed distance to L2 point.  Thus, it is a plausible explanation that this arc-shaped structure is consistent with material leaking out the back door via L2 point.  Such a leakage of material occurs naturally from disks in binaries. The bridge structure emanating from the east side of the SR24N disk can be observed to emanate beyond the size of the outer Roche lobe, indicating that the bridge structure is not attributed to the binary formation between Nb and Nc, but is attributed to the binary formation of the SR24S-N system.

\begin{figure}
\includegraphics[angle=0,scale=0.95]{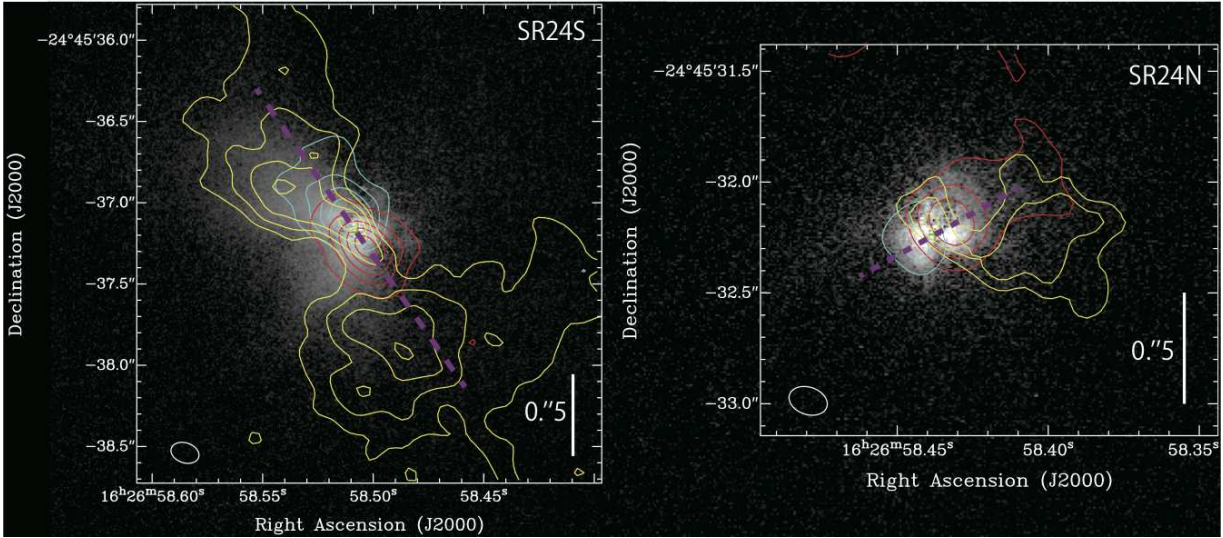}
\caption{Subaru image superimposed on ALMA image \citep{2017ApJ...845...10F} of SR24. Left: $^{12}$CO integrated emission from SR24S overlaid on top of Subaru \textit{PI} scattered light image of SR24S(grey color). Redshifted emission is integrated from 7.3 to 12.4 km s$^{-1}$ (red contours at 27\%, 42\%, 69\%, and 96\% of the peak emission); near-zero-velocity emission is integrated from 2.2 to 4.1 km s$^{-1}$ (yellow contours at 50\%, 65\%, 80\%, and 95\% of the peak emission); blueshifted emission is integrated from -6.0 to 1.6 km s$^{-1}$ (blue contours, same as redshifted contours). Right: $^{12}$CO integrated emission from SR24S overlaid on top of Subaru \textit{PI} scattered light image of SR24S(grey color). Redshifted emission is integrated from 6.6 to 10.5 km s$^{-1}$ (red contours at 20\%, 30\%, and 40\% of the map peak emission located at SR24S); zero-velocity emission is integrated from 5.3 to 6.0 km s$^{-1}$ (yellow contours at 50\% and 60\% of the map peak emission); blueshifted emission is integrated from -0.3 to 2.2 km s$^{-1}$ (blue contours at 50\% and 60\% of the map peak emission).} 
\label{fig13}
\end{figure}

\citet{Schaefer2018} derived the PA and inclination of the SR24Nb-Nc orbit to be 72.0 and 132.1$^{\circ}$, respectively, by calculating the orbit.  As shown in figure \ref{fig13}, \citet{2017ApJ...845...10F} derived the PA and inclination of the secondary SR24N CO disk to be 297$^{\circ}\pm5^{\circ}$ and 121$^{\circ}\pm17^{\circ}$, respectively. Our derived NIR polarization disk PA of 110$^{\circ}$ is roughly consistent with the PA of the CO gas disk. This consistency is because both NIR and CO are tracing surface layers of the disks. Therefore, NIR and CO both traced the circumbinary disk surrounding SR24Nb-Nc. 

The continuum emission detected around SR 24N is unresolved by the ALMA observations at a resolution of 150 [mas] \citep{2017ApJ...845...10F}. As its continuum disk size is much smaller than the SR24Nb-Nc orbit, \citet{Schaefer2018} suggested that the continuum emission is likely from a circumstellar disk surrounding either Nb or Nc and is not from a circumbinary disk around SR 24 Nb-Nc. Based on an orbit calculations by \citet{Schaefer2018}, angular semimajor axis of SR24 N is 181[mas](+83, -30).  By using the estimate from \citet{Artymowicz1994}, namely the outer edge of a circumprimary disc should be truncated at around r = 0.3-0.5 times the semi-major axis, a maximum outer edge of circumstellar disk is 90.5[mas].  This size of disk was not able to be resolved by ALMA observation shown in Fernandez-Lopez et al. 2017.  Therefore, current estimate of circumstellar disk edge agrees with the estimate from \citet{Artymowicz1994}.

\subsection{Asymmetric Disk} 
\label{sect:comp_SMA}

\citet{Andrews10} presented SMA 880~$\mu$m continuum observations of SR24S with a resolution of $0''.37$ and resolved a disk.
Their inset image of the SR24S disk revealed a resolved central emission cavity with an apparent brightness enhancement to the northeast direction. 
According to their model fitting to the visibility at 880~$\mu$m,
the cavity radius is 32 AU (Andrews et al. 2010) or 29 AU (Andrews et al. 2011).

Based on cycle 0 ALMA 0.45 mm continuum observations, \citet{2015A&A...579A.106V} modeled the SR24S disk and derived that its disk PA, inclination, and cavity radius are 20$^{\circ}$, 45$^{\circ}$, and 25 AU, respectively.
They also presented a $^{12}$CO channel map for SR24S, which indicates that the south-west side is moving to the far side, whereas the north-east side is moving to the near side. The zero-moment $^{12}$CO \textit{J} = 6-5 line map in Fig.1 of their paper shows the CO disk extending to the northeast direction. The PA and size of their $^{12}$ CO disk are consistent with the corresponding values of our NIR polarization disk. Similar to the case of SR24N, this consistency is because both NIR and CO are tracing surface layers of the disks. 

\citet{2017ApJ...845...10F} presented 1.3 mm continuum images at a resolution of 0\rlap.{$''$}18 obtained through the ALMA cycle 1 and 2 observations. The ring-shaped disk associated with SR24S is resolved and its semi-major axis, semi-minor axis, PA, and inclination are 0\rlap.{$''$}70$\pm$0\rlap.{$''$}06, 0\rlap.{$''$}50$\pm$0\rlap.{$''$}06, 212$^{\circ}\pm3^{\circ}$, and 44$^{\circ}\pm6^{\circ}$, respectively.  They also derived the PA and inclination of $^{12}$CO disks around both sources as shown in figure \ref{fig13}.  For primary SR24S, the PA and inclination are 218$^{\circ}\pm2^{\circ}$ and 70$^{\circ}\pm5^{\circ}$, respectively. Based on their measurements and analysis, the SR24S disk has its nearest side to the east and the SR24N disk has its nearest side to the north.  They suggest that the SR24S disk rotates in the counterclockwise direction, whereas the SR24N disk rotates in the clockwise direction. 

The cycle 2 ALMA 1.3 mm continuum images of SR24 with a resolution of 0\rlap.{$''$}18 are reported by \citet{2017ApJ...839...99P}.   
The 1.3 mm continuum images of SR24S disk are described by ring-like emission with a central cavity. 
Fitting by \citet{2017ApJ...839...99P} showed that the PA, inclination, and peak radius for the SR24S disk are 24.30$^{\circ}$, 46.31$^{\circ}$, and 0\rlap.{$''$}3, respectively. 
They detected $^{13}$CO and C$^{18}$O($J$=2-1) emission, both of which peaked at the center of the millimeter cavity associated with the SR24S disk.  
Neither continuum nor gas emission from SR24N is detected.  
A potential asymmetric shape on the SR24S disk is inferred from the analysis in the visibility domain. 
Particularly, both the north and south-southeast directions of SR24S have strong emission in contrast with other directions.

Whereas the millimeter cavity around SR24S with a radius of $\sim$0\rlap.{$''$}3 has been resolved by SMA and ALMA, it is not detected in our Subaru image. Thus, SR24S possesses one of the "missing cavities" in NIR scattered light \citep{Dong2012}. Companion-disk interaction combined with dust filtration has been put forward as a likely explanation to such cavities \citep{Zhu2012}; \citep{Dong2015}. Planet-opened gaps can reach a variety of depth depending on the planet mass, disk viscosity, and scale height \citep{Fung2014}. It is possible for gaps to be only modestly depleted in $\sim$micron-sized dust, generally well-coupled to the gas, and not prominent in scattered light. On the other hand, dust filtration \citep{Rice2006} at the outmost gap edge can effectively stop mm-sized dust from entering the gap. Thus, such particles are drained in the inner disk, resulting in a prominent cavity in mm continuum emission. Photoevaporation may also open cavities in disks (e.g., \citep{Alexander2007}). However, a low accretion rate onto the star (\textless 1e-8 M$_\odot$ /year) is expected in this scenario \citep{Owen2012}; \citep{Ercolano2017}, due to its inside-out nature. SR24S has a high accretion rate of 10$^{-7.15}$ M$_\odot$ /year derived from the \textit{Paschen} hydrogen recombination lines\citep{2006A&A...452..245N}, and its cavity is unlikely to be produced by photoevaporation.

Figure \ref{fig7}(b) shows that the radial surface brightness decreases first around 0\rlap.{$''$}5 then stops decreasing till 0\rlap.{$''$}7 along the minor northwest axis.
Figure \ref{fig7}(a) shows that the radial surface brightness decreases first around 0\rlap.{$''$}75 then increases till 0\rlap.{$''$}9 along the major southwest axis.
According to these figure \ref{fig7}(a) and \ref{fig7}(b), both northwest and southwest radial surface brightness shows steeper slope whereas other directions show gradual slope.  The azimuthal direction of this NIR decrement structures is consistent with that observed in submillimeter in \citet{2017ApJ...839...99P}. A possible origin of this asymmetry is discussed in the next subsection.

\subsection{Misaligned inner disk with respect to an outer disk} as an origin of asymmetry 

Recently, \citet{Pinilla2019} reported ALMA band 3 observations at 2.75 mm for the SR24S disk with an angular resolution of 0\rlap.{$''$}11 $\times$ 0\rlap.{$''$}09 and detected an inner disk. They observed that the inner disk emission is likely dominated by dust thermal emission instead of free-free emission. However, it is unclear whether the inner disk is misaligned with respect to the outer disk because the inner disk parameters such as PA, inclination, and gas kinematic information are not derived. 

\citet{Nixon2013, Facchini2013} proposed a mechanism to generate a misaligned disk system: a binary on an inclined orbit with respect to its disk can break the circumbinary disk into inner and outer components, and cause the inner disk to press, resulting in a time-variant mutual inclination between the two disks.

By comparing Subaru NIR and ALMA dust and gas observations with 3D SPH simulation shown in \citet{facchini18}, we interpret that the SR24S disk asymmetry is caused by the misaligned inner disk with respect to the outer disk based on the following two points. 

\begin{figure}
\includegraphics[angle=0,scale=2.0]{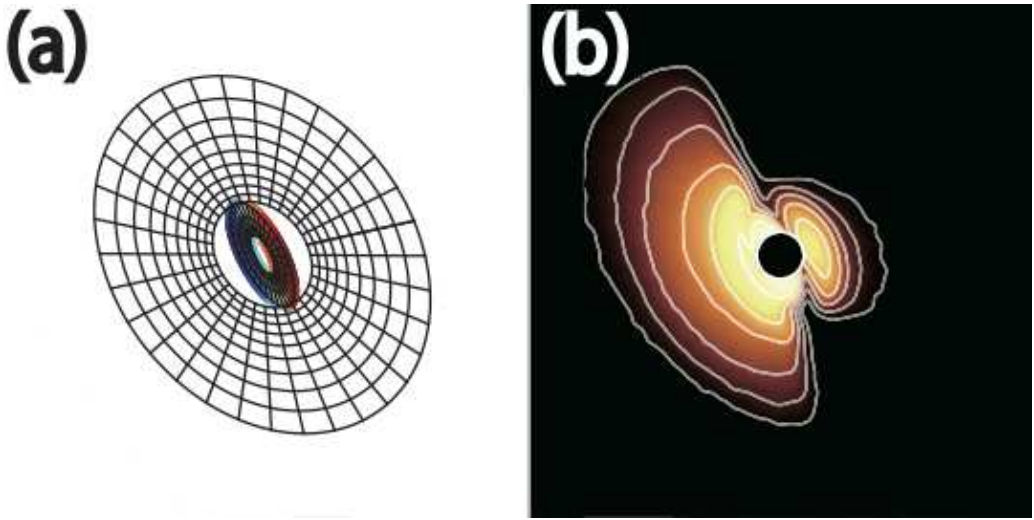}
\caption{3D SPH simulation figures of mis-aligned inner and outer disks cited from \citet{facchini18}. In all panels, the mis-alignment angle between inner and outer disk is $\sim$ 30 $^{\circ}$. All panels are rotated 128 $^{\circ}$ counter-clockwise direction from their original PA in \citet{facchini18} in order to adjust to the PA of SR24 derived from ALMA CO observation.  (a) a schematic of the disc 3D structure cited from Figure 9(b) in \citet{facchini18}. Radial distances are not on scale. (b)scattered light observation at 1.65~$\mu$m of the hydro model cited from Figure 9(j) in \citet{facchini18}. Inclination angle is 45 $^{\circ}$. 
} 
\label{fig4}
\end{figure}

(i) Scattering image: There are two constricted regions toward the north and southwest directions (PA=0 and 225 $^{\circ}$) in the Subaru NIR scattering image. While both sides of the circumprimary disk along the minor axis show mostly a symmetric distribution in the 1.3 mm dust continuum, only the northeast and southern sides of the circumprimary disk in the NIR scattering image are bright. This morphology can be observed in Figure \ref{fig4} (b).

(ii) Dust continuum: The signal-to-noise ratio at both 0.45 and 1.3 mm continuum images of the SR24S disk shows that the west side of the ring has a slightly weaker emission compared with the east side of the ring \citep{2017ApJ...839...99P}. This asymmetry is consistent with Figure 13 (j) in \citet{facchini18}.

As compared in (i) and (ii), the stages of the 3D SPH simulations shown in \citet{facchini18} shared common features with the observed images in the NIR and continuum.  
This consistency between observations and simulation suggests that the observed asymmetry on the circumprimary disk SR24S in NIR scattered light might be affected by the misaligned inner disk with respect to the outer disk.  

A comparison between 3D SPH simulation by \citet{facchini18} and observations also provides constraints on the inclination of the inner disk. We compared Figure 8 for the $\xi$ = 74 $^{\circ}$ case and Figure 9 for the $\xi$ = 30 $^{\circ}$ case in \cite{facchini18}). ($\xi$ denotes the misalignment angle between the inner and outer disks). In particular, (i), (j), (k), and (l) panels in both Figures 8 and 9, which have an outer disk inclination of 45 $^{\circ}$, are compared because previous submillimeter dust continuum observations revealed that the SR24S outer disk has an inclination of approximately 45 $^{\circ}$.  The outer disk shows a relatively axisymmetric structure, with two azimuthal regions of lower surface brightness for the $\xi$ = 74 $^{\circ}$ cases. For the $\xi$ = 30 $^{\circ}$ cases, in contrast, the outer disk show a relatively non-axismmetric structure, with one side being much brighter than the other. In addition, there are relatively less clearer signatures of pairs of azimuthal intensity decrements at near-symmetric locations in contrast with the $\xi$ = 74 $^{\circ}$ cases. As our NIR image has a similar asymmetric structure, the inclination of the inner disk can be constrained to close to $\xi$ = 30 $^{\circ}$ cases in contrast with $\xi$ = 74 $^{\circ}$ cases.

Finally, the leading formation mechanism of the misaligned inner disk with respect to the outer disk around SR24S is discussed here. As introduced in Section 1, there are mainly three promising mechanisms that theoretically claimed to address the origin of misalignment between an inner and an outer disk:
1) the rotation axis of the disk system is misaligned with respect to the magnetic field direction; 2) anisotropic accretion of gas with different rotational axes; and  
3) a misaligned massive planet with respect to an outer disk tilting an inner disk.
To discuss the leading formation mechanism of the misaligned inner disk with respect to the outer disks in the SR24S case and provide constraints to these mechanisms, we list some observational results below.

As discussed above, the misalignment angle between the inner and outer disks can be constrained to close to $\xi$ = 30 $^{\circ}$ in contrast with $\xi$ = 74 $^{\circ}$.  The third mechanism starts from a small misalignment angle between an inner and outer disk and eventually produces a large misalignment angle, whereas the first mechanism can only produce small misaligned angles. Therefore, the first mechanism can be ruled out. Although no direct imaging observations have detected a companion inside the SR24S cavity so far, the third mechanism triggered by an undetected massive companion embedded in the cavity could possibly tilt the inner disk of SR24S. Subsequently, the second mechanism cannot be ruled out because the mass accretion rate of SR24S and SR24N is 10 $^{-7.15}$ and 10 $^{-6.90}$ ~\textit{M}$_\odot$/year derived from the \textit{Paschen} hydrogen recombination lines\citep{2006A&A...452..245N}, respectively. In addition, the circumprimary disk around SR24S has a bridge and spiral arm. According to the numerical simulation in \citet{Mayama10}, fresh material streams along the spiral arm in which gas is replenished from a circummultiple reservoir and the bridge corresponds to gas flow and a shock wave caused by the collision of gas rotating around the primary and secondary stars. These structures, in particular the bridge, might contribute to tilting of the outer disk around SR24S. This is because the bridge is physically connecting the two circumprimary and circumsecondary disks, which are strongly misaligned with one another. While it is difficult to provide further constraints to the origin of misalignment between the inner and outer disk with the currently available data, these mechanisms can be revealed using very-high-resolution observations such as ALMA in the future.

\subsection{Binarity of SR24S}

The similarities between SR24S and HD142527 suggest the presence of a relatively massive companion. \citet{Price2018} and \citet{Lacour2016} demonstrated that the presence of such a companion can address various structures observed in HD142527 disk including cavity, horseshoe, or so on. Therefore, we discuss the possibility that SR24 S may have an unseen companion.

As introduced in section 1, spectral type L2 and a luminosity of 12.9~\textit{L}$_\odot$ of the SR24S is adopted here from \citet{Greene1995}.  Although the extinction correction of Av=13.7 mag is large, this would not change the conclusion.  It is because that would not move SR24S horizontally, but vertically on the HR diagram. Figure 5 shows that SR24S is plotted along with the pre-main sequence evolution tracks derived in \citet{Tognelli2011}.  This figure shows the mass is slightly larger than 2.0~\textit{M}$_\odot$.

Since PMS star properties in both \citet{Greene1995} and Pecaut \& Mamajek provide T$_{eff}$ = 5000 $K$ and 5040 $K$, respectively, log $t$ = 3.70 is a reliable parameter.  Mass of SR24S is derived as 2.0~\textit{M}$_\odot$ in \citet{Greene1995}.

Consequently, taken all the uncertainties into account, it would be hard for SR24S to have an equal mass binary star. This is because the combined light would be cooler than a K2 star if SR24S and its binary have both 0.9~\textit{M}$_\odot$ for example.  However, it is possible for SR24S to have a companion which have less than 0.4~\textit{M}$_\odot$ as the smaller mass companion would be from 1/10 to 1/20 the luminosity of the more massive star, SR24S.

Furthermore, \citet{Pinilla2016} and \citet{Pinilla2019} used ALMA observation data and demonstrated that a massive planet (\textless 5~\textit{M}$_{jup}$) would be present in the cavity of SR24S while they excluded the possibility of existence of more massive planets ($\gtrsim$ 5~\textit{M}$_{jup}$) in the cavity of SR24S. The misalignment between inner and outer disk surrounding SR24S discussed in 4.3. might be attributed to this embedded massive companion. According to 3D numerical simulations by \citet{Nealon2018}, for a planet massive enough to carve a gap, a disk is separated into two components and the gas interior and exterior to the planet orbit evolve separately, forming an inner and outer disk. Due to the inclination of the planet, a warp develops across the planet orbit such that there is a relative tilt and twist between these disks.

\section{Summary}

 We have conducted high-resolution $H$-band polarimetric imaging observations of the enigmatic SR24 triple system. The main conclusions are as follows:

1. The circumprimary disk associated with SR24S is resolved and has elongated features both to the northeast and southeast directions. The PA and radius of the NIR polarization disk around SR24S are 55$^{\circ}$ and 1\rlap.{$''$}2, respectively. The PA and size of the $^{12}$CO disk are consistent with the corresponding values of our NIR polarization disk. As the stages of the 3D SPH simulations shared common features with the observed images in the NIR, continuum, and $^{12}$CO, this consistency suggests that the observed asymmetry on the circumprimary disk might be due to the misaligned inner disk with respect to the outer disk.

2. The circumsecondary disk associated with SR24N is resolved and has elongated features in the east-west direction. The PA and radius of NIR polarization disk around SR24N are 110$^{\circ}$ and 0\rlap.{$''$}3, respectively. The sizes and PAs derived from NIR polarization and $^{12}$CO gas observations are consistent with each other. As the radius of the polarization disk around SR24N measured to be 0."3 is roughly consistent with the computed size of the outer Roche lobe, it is natural to interpret the polarization disk around SR24N detected with HiCIAO as a circumbinary disk surrounding the SR24Nb-Nc system.

3. In the radial direction, the surface brightness of SR24S and SR24N decreases as $r$$^{-1.5}$ from 0\rlap.{$''$}2 to 1\rlap.{$''$}0 and $r$$^{-2.1}$ from 0\rlap.{$''$}1 to 0\rlap.{$''$}3, respectively.
The azimuth radial surface brightness of SR24N has a steeper profile than that of SR24S. Our observations also show that the SR24S disk is more spatially extended than the SR24N disk. 

4. As an overall morphology, the circumprimary disk around SR24S shows strong asymmetry, whereas the circumsecondary disk around SR24N shows relatively strong symmetry. Both the circumprimary and circumsecondary disks show similar structures as the $^{12}$CO gas disk in terms of size and elongation direction. This consistency is because both NIR and $^{12}$CO are tracing surface layers of the flared disks. Our NIR observations confirm the previous claim made through 0\rlap.~{$''$}2 submillimeter observations that the circumprimary disk is misaligned with respect to the circumsecondary disk.



\acknowledgments

We thank the telescope staff and operators of the Subaru Telescope for their assistance.  
We sincerely thank the referee for giving us all these very useful and constructive suggestions which greatly improves the paper.
M.T. is partially supported by JSPS KAKENHI Grant Numbers 18H05442 and 15H02063. 
This work was supported in part by The Graduate University for Advanced Studies, SOKENDAI and JSPS KAKENHI grant Numbers 25800107.

\clearpage



\begin{figure}
\includegraphics[angle=0,scale=0.8]{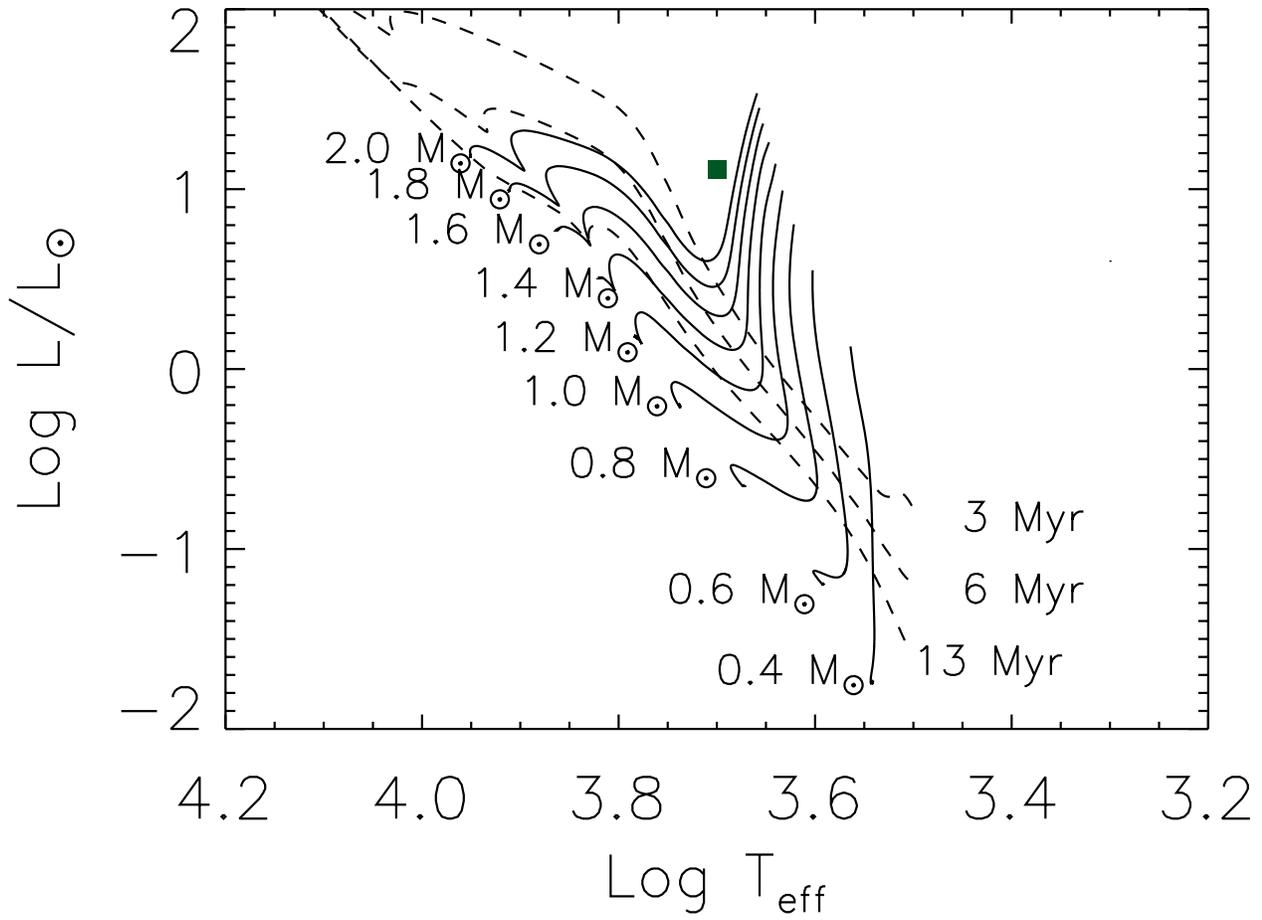}
\caption{SR24S primary star plotted as green square along with the PMS evolution tracks of \citet{Tognelli2011}}
\label{fig5}
\end{figure}



\end{document}